# Extraction of Zero-Point Energy from the Vacuum: Assessment of Stochastic Electrodynamics-Based Approach as Compared to Other Methods

**Garret Moddel * and Olga Dmitriyeva**

Department of Electrical, Computer, and Energy Engineering, University of Colorado, Boulder, CO 80309-0425, USA; moddel@colorado.edu
* Correspondence: moddel@colorado.edu; Tel.: +1-303-492-1889

**Abstract:** In research articles and patents several methods have been proposed for the extraction of zero-point energy from the vacuum. None of the proposals have been reliably demonstrated, yet they remain largely unchallenged. In this paper the underlying thermodynamics principles of equilibrium, detailed balance, and conservation laws are presented for zero-point energy extraction. The proposed methods are separated into three classes: nonlinear processing of the zero-point field, mechanical extraction using Casimir cavities, and the pumping of atoms through Casimir cavities. The first two approaches are shown to violate thermodynamics principles, and therefore appear not to be feasible, no matter how innovative their execution. The third approach, based upon stochastic electrodynamics, does not appear to violate these principles, but may face other obstacles. Initial experimental results are tantalizing but, given the lower than expected power output, inconclusive.

---

## 1. Introduction

We present results from a set of experiments in which we attempted to extract quantum vacuum energy from a gas flow system based upon stochastic electrodynamics principles. To put this method in context we first outline other methods that have been proposed for harvesting this energy. The methods are divided into three classes, and the underlying principle of operation for each is assessed in terms of which would violate fundamental thermodynamics principles.

Physical effects resulting from zero-point energy (ZPE) are well established ([1], Chapter 3). This has led to several proposals and reviews discussing the extraction of ZPE to use as a power source [2–9]. The ZPE extraction methods usually involve ZPE in the form of electromagnetic zero-point fields (ZPFs). The energy density of these ZPE vacuum fluctuations is ([1], p. 11):

$$\rho(h\nu) = \frac{8\pi\nu^2}{c^3}\left(\frac{h\nu}{\exp(h\nu/kT)-1} + \frac{h\nu}{2}\right), \tag{1}$$

where ν is the frequency, and $k$ is Boltzmann's constant. The first term in the large brackets describes Planck radiation from a black body at temperature $T$. At room temperature, at frequencies above 7 THz, the energy density is dominated by the second, temperature-independent term, which is due to zero-point energy. For high frequencies this energy density is large, but just how large depends upon the frequency at which the spectrum cuts off, a matter that is not yet resolved.

Two physical manifestations of the ZPF that will be discussed in this paper are zero-point noise fluctuations and the force between Casimir cavity plates. The available noise power in a resistance $R$ per unit bandwidth is [10]

$$\frac{\Delta V^2}{4R} = \frac{h\nu}{\exp(h\nu/kT)-1} + \frac{h\nu}{2}, \tag{2}$$

The first term on the right-hand side is the thermal noise, which is approximated at low frequencies by the familiar Johnson noise formula. The second term, usually called quantum noise, is due to zero-point fluctuations. This physical manifestation of the ZPF dominates the noise at low temperatures and high frequencies.

A second physical manifestation is evident with a Casimir cavity, which consists of two closely-spaced, parallel reflecting plates [11]. Because the tangential electric field is reduced (to zero for a perfect conductor) at the boundaries, limits are placed on the ZPF modes between the plates, and those modes having wavelengths longer than twice the gap spacing are suppressed to a degree determined by the plate reflectivity. This suppression is not imposed on the ZPF on the exterior side of the plats. The radiation pressure exerted on the exterior side of the plates is larger than on the interior, with the result that the plates are pushed together. The resulting attractive force between the plates is ([1], p. 58):

$$F(d) = \frac{\pi^2 \hbar c}{240 d^4}, \tag{3}$$

where $d$ is the gap spacing. For this force to be measurable with the currently available experimental techniques, $d$ must be less than 1 μm.

The benefits of tapping ZPE from the vacuum would be tremendous. Assuming even a conservative cutoff frequency in Equation (1), if just a small fraction of this energy were available for extraction, the vacuum could supply sufficient power to meet all of our needs for the foreseeable future. Cole and Puthoff [12] have shown that extracting energy from the vacuum would not, in principle, violate the second law of thermodynamics, but that is not equivalent to stating that extraction is feasible, nor do they attempt to describe how it could be accomplished. There is no verifiable evidence that any proposed method works [13], and there has been a continuing debate as to whether ZPE is fundamentally extractable [14].

In this article, we assess different methods that have been proposed to extract usable ZPE. We do not examine proposed methods to use ZPE forces as a means to enhance or catalyze the extraction of energy from other sources, such as chemical [15] or nuclear energy. We separate the different vacuum energy extraction approaches into three classes: nonlinear extraction, mechanical extraction, and pumping of gas. We analyze each to see if the underlying principles of operation are consistent with known equilibrium thermodynamics principles, and then draw conclusions about the feasibility of the ZPE extraction. Finally, we describe a method based on a stochastic electrodynamics picture of ZPE and present initial experimental results using the approach.

## 2. Analysis

### *2.1. Nonlinear Processing of the Zero-Point Field*

#### 2.1.1. Rectification of Zero-Point Fluctuations in a Diode

Several suggested approaches to extracting energy from the vacuum involve nonlinear processing of the ZPF. One particular nonlinear process is electrical rectification, in which an alternating (AC) waveform is transformed into a direct (DC) one.

Valone [9] describes the electrical noise in resistors and diodes that results from zero-point fluctuations. He discusses the use of diodes to extract power from these ambient fluctuations, and compares this to diodes used for thermal energy conversion. For example, in thermophotovoltaics radiation from a heated emitter is converted to electricity. In Valone's case, however, the source is under ambient conditions. In support of Valone's approach, a straightforward analysis of diode rectification in the presence of thermal noise does appear to produce a rectified output from that noise unless a compensating current is added to the system [16]. Valone is particularly interested in the use of zero-bias diodes for zero-point energy harvesting, so as to rectify the ambient fluctuations without having to supply power in providing a voltage bias to the diodes.

This nonlinear extraction represents a sort of Maxwell's demon [17]. In 1871 Maxwell developed a thought experiment in which a tiny demon operates a trapdoor to separate gas in equilibrium into two compartments, one holding more energetic molecules and the other holding less energetic ones. Once separated, the resulting temperature difference could be used to do work. This is a sort of nonlinear processing, in which the system, consisting of the demon and the compartments, operates differently on a molecule depending upon its thermal energy. In the nearly fifteen decades since its creation, innovative variations on the original demon have been proposed and then found to be invalid. Despite the best efforts of Maxwell's demon and his scrutineers [18,19] there still is no corroborated experimental evidence for the demon's viability [20], and for the current analyses I will assume that he cannot assist us in extracting energy from a system in equilibrium.

In the absence of such a demon, thermal noise fluctuations cannot be extracted without the expenditure of additional energy, because such fluctuations are in a state of thermal equilibrium with their surroundings [21]. These thermal fluctuations are described by the first term on the right-hand side of Equation (2). In equilibrium, the second law of thermodynamics applies and no system can extract power continuously. All processes in such a system are thermodynamically reversible. A detailed balance description of the kinetics of such a situation was developed by Einstein to explain the relationship between the emission and absorption spectra of atoms [22], and discussed as a manifestation of equilibrium by Bridgman [23].

If the equilibrium is altered, for example by the addition of a non-equilibrium radiation field, then the detailed balance is replaced by a less restrictive steady state condition and it becomes possible to extract power. The difference between detailed balance and steady state is illustrated with the three-state system shown in Figure 1. Each arrow represents a unit of energy flux. In the steady-state case shown in Figure 1a, the total flux into any state equals the total flux out of it. Under equilibrium, however, a more restrictive detailed balance must be observed, in which the flux between any pair of states must be balanced. This is depicted in Figure 1b.

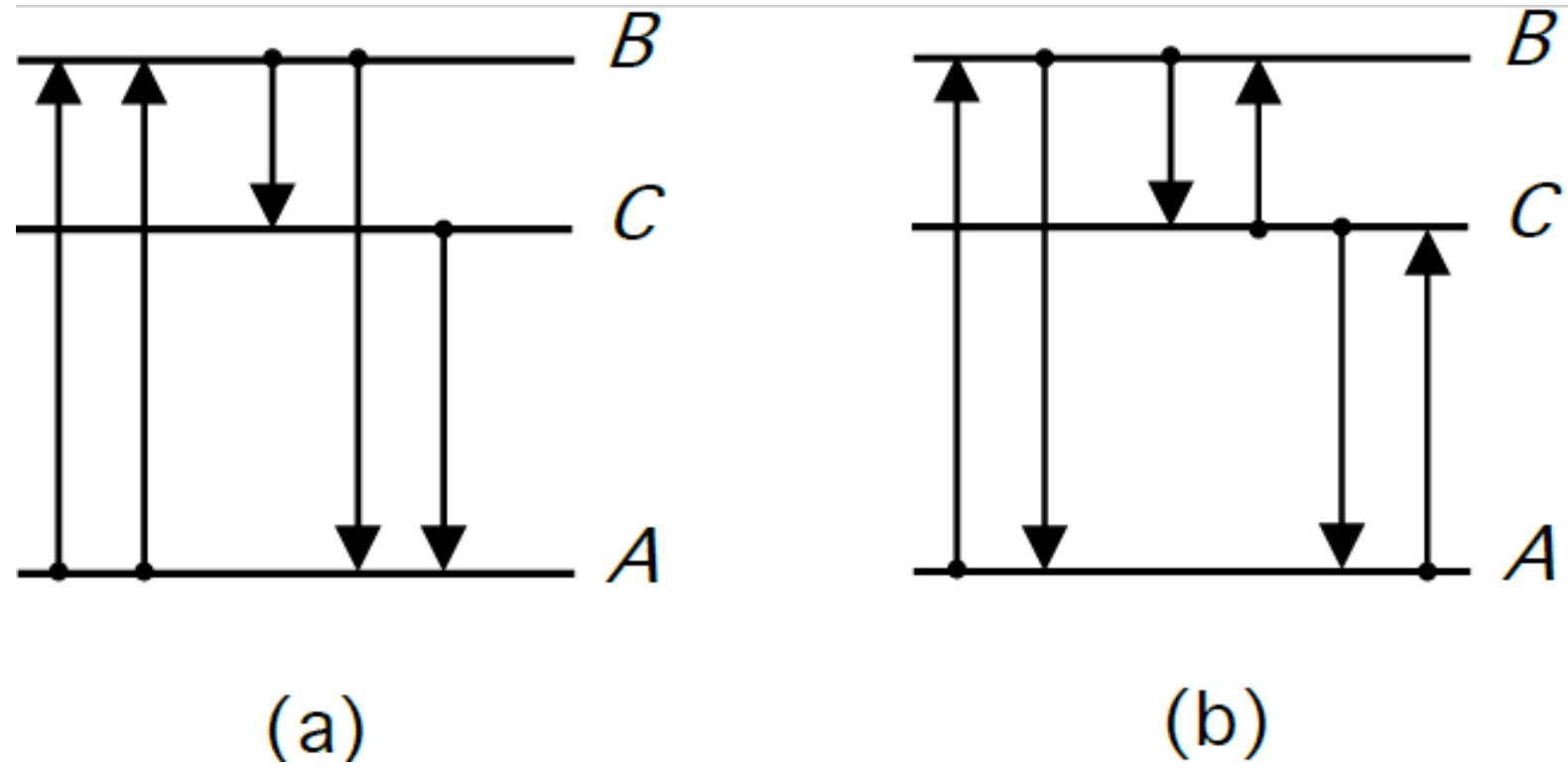

**Figure 1.** Illustration of detailed balance. In this three-state system each arrow represents one unit of energy flux. The system in (**a**) is in steady state, such that the total flux into each state equals the total flux out of it; In system (**b**) not only does the steady-state condition apply, but the more restrictive detailed balance also applies, in which the flux between each pair of states is balanced.

This concept of detailed balance can be applied to the extraction of thermal noise from a resistor at ambient (equilibrium) temperature. To optimally transfer power from a source, in this case the noisy resistor, to a load the load resistance should be adjusted to match that of the source. In that case, the load generates an equal noise power to that of the source, and an equal power is transferred from the load to the source as was transferred from the source to the load. Because of this detailed balance, no net power can be extracted from a noisy resistor.

To analyze the case of extracting energy from thermal noise fluctuations in a diode, consider the energy band diagram for a diode shown in Figure 2, where transitions among three different states are shown. For simplicity, five other pairs of transitions are not shown and are assumed to have negligible rates. (Not shown are (i) direct transitions between the p-type region valence band and the n-type region conduction band; (ii) transport of valence-band charge carriers (holes) through the junction region; (iii) generation and recombination in the n-type region; (iv) direct transitions between the n-type region valence band and the p-type region conduction band; and (v) generation and recombination in the junction region. A completely parallel set of processes could be added for these transitions, and would not change the physical principles involved, or the conclusions drawn.)

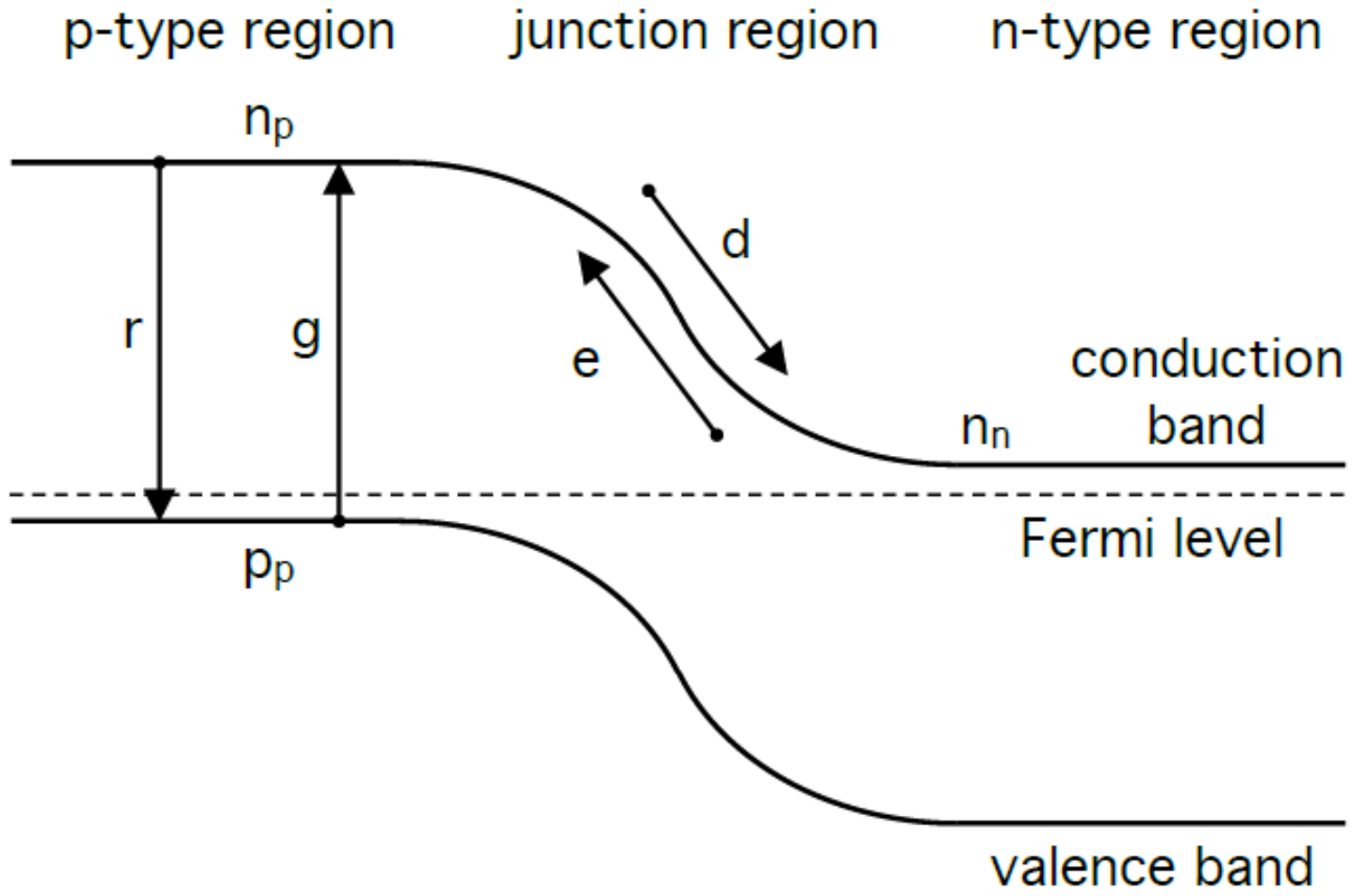

**Figure 2.** Diode energy band diagram. Shown are electron transitions between the conduction and valence bands in the p-type region, corresponding to generation rate *g*, and recombination rate *r*. Also shown are electron transitions between n-type and p-type conduction band states, corresponding to excitation rate *e*, and drift rate *d*. This diagram is used in the text to illustrate photovoltaic carrier collection, rectification of thermal fluctuations, and also rectification of zero-point energy fluctuations as proposed by Valone [9].

First consider the case of photovoltaic power generation. If the diode operates as a solar cell, light absorbed in the p-type region generates electron-hole pairs, promoting electrons to the conduction band at a rate $g$ that depends upon the light intensity and other factors. The photogenerated electrons diffuse to the junction region, where the built-in electric field causes them to drift across the junction to the n-type region at rate $d$. The recombination rate, $r$, and the excitation rate, $e$, are also shown. Because $g \gg r$ and $d \gg e$ under solar illumination, i.e., the system is far from equilibrium, there is a net flow of electrons to the n-type region, where they are collected to provide power.

The diode would operate in much the same way for rectifying thermal fluctuations under equilibrium, except that now $g$ would represent the thermal generation rate. In this case, however, the generation rate and drift rate across the junction would be much smaller than under solar illumination. Under thermal equilibrium and in the absence of a Maxwell's demon to influence one of the processes, a detailed balance is strictly observed, such that $g = r$ and $d = e$. The second law of thermodynamics does not allow for power generation.

Valone's proposal [9] makes use of power generation in a diode from ZPE fluctuations described by the second term on the right-hand side of Equation (2). Whether this is feasible becomes a question of whether the zero-point energy in a diode is in a state of true equilibrium with its surroundings. It has been generally accepted that the "'vacuum' should be considered to be a state of thermal equilibrium at the temperature of $T = 0$" [12]. Recently, using the principle of maximal entropy, Dannon has shown explicitly that zero-point energy does, in fact, represent a state of thermodynamic equilibrium [24]. Therefore, it is clear that the detailed balance argument presented above for the case of thermal fluctuations also applies to ambient zero-point energy fluctuations, and a diode cannot rectify these fluctuations to obtain power.

2.1.2. Harvesting of Vacuum Fluctuations Using a Down-Converter and Antenna-Coupled Rectifier

A somewhat different approach to the nonlinear processing of the ZPF for extracting usable power would be to use an antenna, diode and battery. The radiation is received by the antenna, rectified by the diode, and the resulting DC power charges the battery. In the microwave engineering domain, this rectifying antenna is known as a rectenna [25]. Because of its $\nu^3$ dependence, shown in Equation (1), the ZPF power density at microwave frequencies is too low to provide practical power. Therefore, to obtain practical levels of power, a rectenna must operate at higher frequencies, such as those of visible light or even higher. There are diodes that operate at petahertz frequencies. One example is a graphene geometric diode [26] but the rectification power efficiency of optical rectennas at such high frequencies is generally low [27]. The first question about ZPF rectification is how it can be made practical. The second, and more important question here, is whether this is feasible from fundamental considerations. I address these in turn.

A method to extract ZPE is proposed in a 1996 patent by Mead and Nachamkin [4], which describes an invention to produce lower beat frequency from high-frequency ZPF to make it more practical to rectify. The invention includes resonant microscopic spheres that intercept ambient ZPF and build its intensity at their resonant frequency. The high-intensity oscillation induces interactions between two spheres of different size such that a lower beat-frequency radiation is emitted from them. This lower beat-frequency radiation is said to be then absorbed by an antenna and rectified to provide DC electrical power. (Note that the beat frequency is not frequency down-conversion, which requires a nonlinear mixer. The down-conversion occurs only after the signals encounter the diode, and so the antenna cannot actually pick up the short-wavelength ZPF, and the diode cannot rectify the high-frequency ZPF.) The invention is depicted in Figure 3.

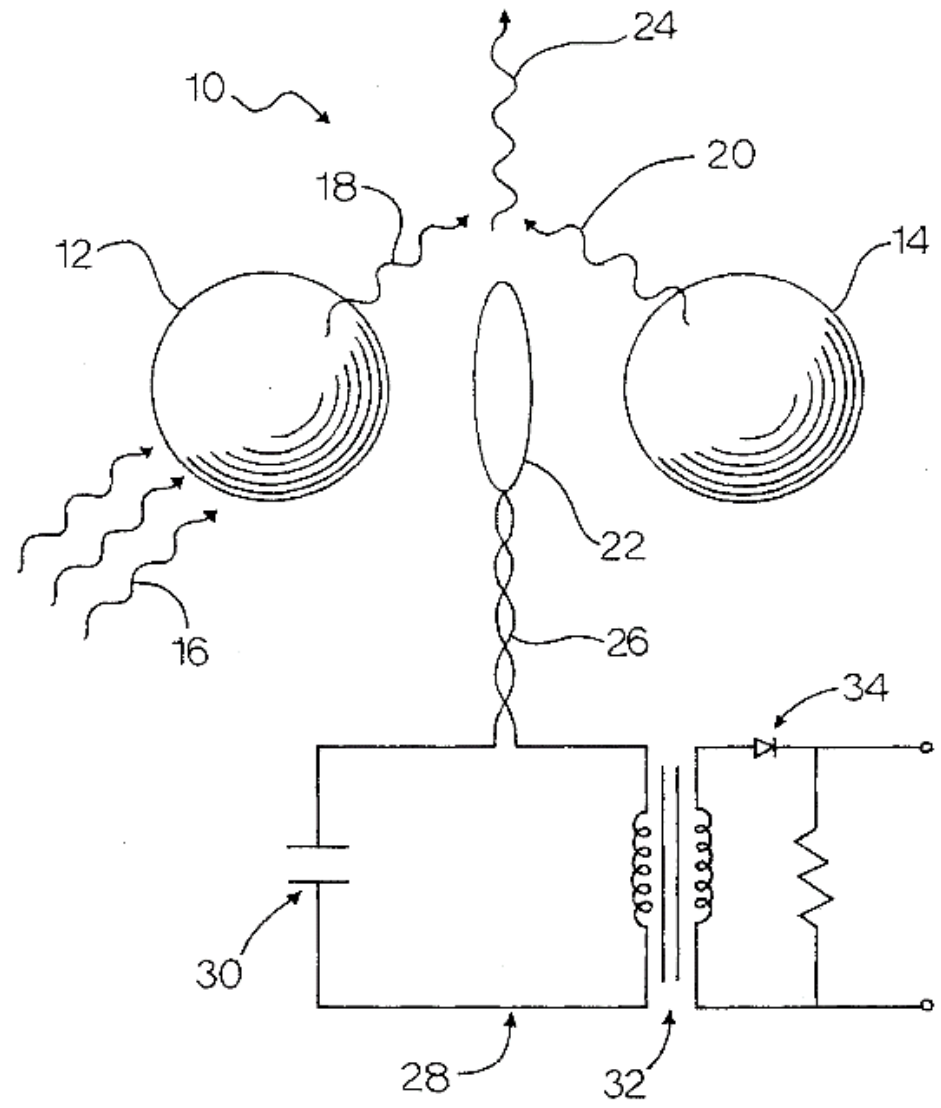

**Figure 3.** Mead and Nachamkin's invention [4] for producing a beat frequency from zero-point field radiation, and collecting, and rectifying it. The non-identical microscopic resonant spheres interact with ambient zero-point fields to produce radiation at a beat frequency. This radiation is absorbed in the loop antenna and rectified in the circuit.

Mead and Nachamkin's approach can be broken down into three steps:

(a) Producing lower beat frequency radiation from the ambient ZPF;
(b) Collecting the beat-frequency radiation at the diode by the antenna;
(c) Rectification of the concentrated radiation by the diode.

Step (a) is a process akin to that performed by the diode described in the previous section, except that in the current case the ZPF produces an intermediate beat frequency whereas in the previous case the ZPF fluctuations in a diode were down-converted to DC. Therefore, this step operating with incoming radiation under equilibrium must observe a detailed balance of rates. Regarding steps (b) and (c), under equilibrium a source, antenna and load are in detailed balance, such that the power received by the antenna from the source and transferred to a load is equal to the power transmitted back to the source [28].

If steps (a) and (b) could provide a greater-than-equilibrium concentration of power to the diode, then the diode in step (c) would no longer be operating under equilibrium. When driven far from equilibrium the harvesting efficiency would be limited to the Carnot efficiency in a quantum heat engine, unless the extra energy somehow created some coherence [29]. However, as argued above, the concentration of power at the diode cannot occur under equilibrium. In summary, when applied to the harvesting of vacuum fluctuations each of the three steps in the ZPF down-converter system is subject to a detailed balance of rates, and therefore, the system cannot provide power.

#### 2.1.3. Nonlinear Processing of Background Fields in Nature

If nonlinear processing of ZPFs were sufficient to extract energy, one would expect to see the consequences throughout nature. Naturally occurring nonlinear inorganic and organic materials would down-convert the ZPF, for example, to the infrared. The result would be constant warmth emanating from these nonlinear materials. Such down-conversion may exist, but through detailed balance there must be an equal flux of energy from the infrared to higher-frequency background ZPF.

For ZPFs to provide usable power, an additional element must be added beyond those providing nonlinear processing of ambient fields.

### *2.2. Mechanical Extraction Using Casimir Cavities*

The attractive force between two closely spaced conducting, i.e., reflecting, plates of a Casimr cavity was predicted by Casimir in 1948 [11], and is given by Equation (3). This attractive force was later shown to apply also to closely spaced dielectric plates [30], and becomes repulsive under certain conditions [31]. The potential energy associated with the Casimir force is considered next as a source of extractable energy [5,6].

The simplest way to extract energy from Casimir cavities would be to release the closely spaced plates so that they could accelerate together. In this way, the potential energy of the plate separation would be converted to kinetic energy. When the plates hit each other, their kinetic energy would be turned into heat. The Casimir cavity potential would be extracted, albeit into high-entropy thermal energy. If this energy conversion could be carried out as a cyclic process, electrical power obtained from this heat would be subject to the limitations of the Carnot efficiency:

$$\eta_{\max} = \frac{T_h - T_c}{T_h} \tag{4}$$

where $T_h$ is the temperature of hot source and $T_c$ is the temperature of the cold sink.

#### 2.2.1. Energy Exchange between Casimir Plates and an Electrical Power Supply

In 1984, Forward described a different concept [2] for extracting energy from the mechanical motion of Casimir plates, one that maintains the low entropy of the Casimir cavity's potential energy through the extraction process. A coiled Casimir plate is shown in Figure 4. The attractive Casimir force between spaced-apart coils of the Casimir plates is nearly balanced by the injection of electric charge from an external power supply causing the plates to repel each other. As the plates move together due to the attractive Casimir force, they do work on the repulsive charge, resulting in a charge flow and transfer of energy to the power supply. In this way, the coming together of the Casimir plates provides usable energy, and maintains the low entropy of the original attractive potential energy. Forward made no attempt to show how this would provide continuous power, since once the plates came together all the available potential energy would be used up.

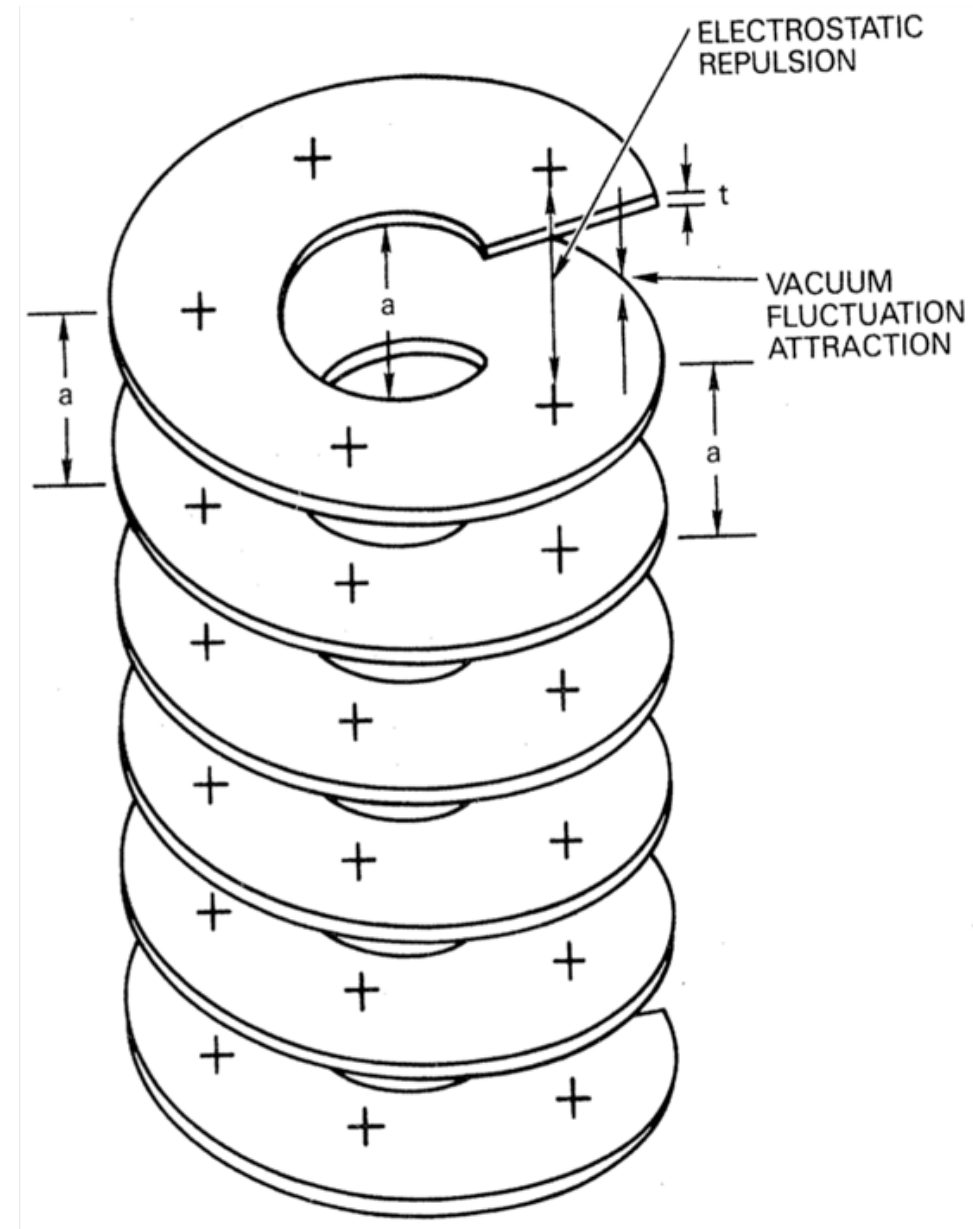

**Figure 4.** Slinky-like coiled Casimir cavity, as conceptualized by Forward [2]. He used this device concept to demonstrate how one might convert the vacuum fluctuation potential energy from Casimir attraction to electrical energy. As the plates approach each other the repulsion of positive like-charges results in a current that charges up an external power supply.

2.2.2. Cyclic Power Extraction from Casimir Cavity Oscillations

In Forward's concept, described in the previous section, energy is extracted from the Casimir force until the plates have come together, but continuous power extraction would require a cyclic process. In a series of publications, Pinto proposed an engine for the extraction of mechanical energy from Casimir cavities [6]. His concept makes use of switchable Casimir cavity mirrors. A schematic depiction of the process is shown in Figure 5. In step (a) the Casimir cavity plates are allowed to move together in response to their attraction, and the reduction in potential energy is extracted (for example, by the Forward method). In step (b) one of the plates is altered to change its reflectivity. Because of the altered state, the attractive Casimir force is reduced or reversed and the plates can then be separated using less energy than was extracted when they came together, as depicted in step (c). After they are separated, the plates are restored to their original state, and the cycle is repeated. Puthoff analyzed a system of switchable Casimir cavity mirrors and calculated the potential power that could be produced by Casimir plates as a function of vibration frequency and mass [32].

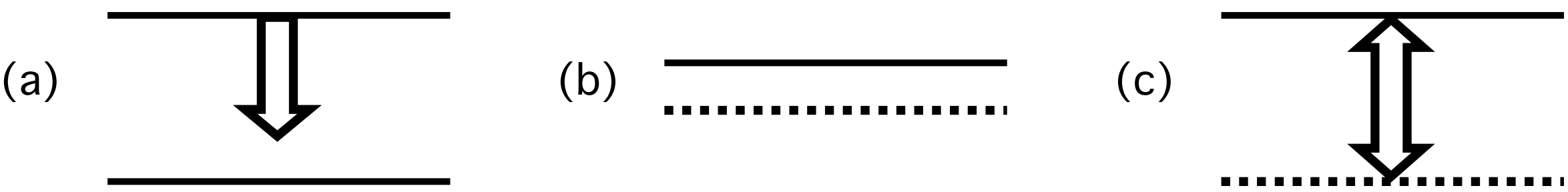

**Figure 5.** Casimir cavity engine for the cyclic extraction of vacuum energy, similar to system proposed by Pinto [6]. In step (**a**) the Casimir plates move together in response to Casimir attraction, producing energy that is extracted; In step (**b**) the lower plate is altered to reduce its reflectivity and hence reduce the Casimir attraction; In step (**c**) the plates are pulled apart, using less energy than that which was obtained in step (**a**), and then the cycle is repeated.

Pinto's approach cannot work if the Casimir force is conservative. If so, no matter what process were to be used in separating the plates, it would require at least as much energy as had been extracted by their coming together. For example, in step (b) shown in Figure 5, electrical charge might be drained from at least one of the plates to modify its reflective property. This would reduce the Casimir attraction and allow the plates to be pulled apart with minimal force, after which charge would be injected back into the plates to reestablish the Casimir attraction. For a conservative force, the minimum energy required for this draining and injecting back of the electrical charge is the energy that could be obtained from the attractive force of the plates moving together. Alternatively, exposing a plate to hydrogen may be used to change its reflectivity [33] and hence the attraction between plates. If the force is conservative then the hydrogenation/de-hydrogenation cycle would require at least as much energy as could be extracted from the Casimir-plate attraction, and the system cycle could not produce power. A similar situation to that of the Casimir force exists with standard electric forces, which clearly are conservative. The electric attraction induced by opposite charge on two capacitor plates cannot produce cyclic power.

In one analysis, a Carnot-like cycle was used to show that the Casimir force did not appear to be conservative [6], so that it would be possible to extract cyclic power. However, from an analysis of each of the steps in the cycle, Scandurra found that the Casimir force is conservative after all, consistent with the general consensus, and that the method cannot produce power in a continuous cycle [34]. In a recent publication, Pinto has supported this conclusion and its implications [35].

Generalizing from the conservative nature of the Casimir force, it appears that any attempt to obtain net power in a cyclic fashion from changing the spacing of Casimir cavity plates cannot work. In a different sort of mechanical extraction, it may be possible to use vacuum fluctuations as nanoscale hammers to heat surfaces [36], or even to account for the expansion of the Universe via parametric resonance [37].

### *2.3. Pumping Atoms through Casimir Cavities*

#### 2.3.1. Zero-Point Energy Ground State and Casimir Cavities

There is a fundamental difference between the equilibrium state for heat and for ZPE. It is well understood that one cannot make use of thermal fluctuations under equilibrium conditions. To use the heat, there must be a temperature difference to promote a heat flow to obtain work, as reflected in the Carnot efficiency of Equation (4). We cannot maintain a permanent temperature difference between a hot source and a cold sink in thermal contact with each other, without expending energy, of course.

Similarly, without differences in some characteristic of ZPE in one region as compared to another, it is difficult to understand what could drive a flow of ZPE to allow its extraction. If the ZPE represented the universal ground state, we could not make use of ZPE differences to do work. But the entropy and energy of ZPE are geometry dependent [38], and are a function of the boundary conditions [39]. In this way ZPE fluctuations differ fundamentally from thermal fluctuations. Inside a Casimir cavity the ZPF density is different than outside, a difference that is established as a result of the different boundary conditions inside and out. A particular state of thermal or chemical equilibrium can be characterized by a temperature or chemical potential, respectively. For an ideal Casimir cavity having perfectly reflecting surfaces it is possible to define a characteristic temperature that describes the state of equilibrium for zero-point energy and which depends only on cavity spacing [33]. In a real system, however, no such parameter exists because the state is determined by boundary conditions in addition to cavity spacing [40], such as the cavity reflectivity as a function of wavelength, spacing uniformity, and general shape.

The next approach to extracting power from vacuum fluctuations makes use of the step in the ZPE ground state at the entrance to Casimir cavities. According to stochastic electrodynamics (SED), the energy of classical electron orbits in atoms is determined by a balance of emission and absorption of vacuum energy [41,42]. By this view of the atom, electrons emit a continuous stream of Larmor radiation as a result of the acceleration they experience in their orbits. As the electrons release energy their orbits would spin down were it not for absorption of vacuum energy from the ZPF. This balancing of emission and absorption has been modeled and shown to yield the correct Bohr radius in hydrogen [43–45]. Accordingly, based on unpublished suggestions by B. Haisch and H.E. Puthoff, the orbital energies of atoms inside Casimir cavities should be shifted if the cavity spacing blocks the ZPF required to support a particular atomic orbital. However, under particular simplifying assumptions this shift is not predicted [46]. This reduction of orbital energy inside Casimir cavities is not associated with a reduced temperature, as the entire system is in thermal equilibrium. Unlike with the Lamb shift, the orbital energy is not just modified by the ZPF but is directly supported by it, and so the reduction in energy inside a Casimir cavity is expected to be much larger than the change in the Lamb shift as predicted by cavity quantum electrodynamics [47].

We do not attempt to assess whether SED is valid, which is well beyond the scope of this investigation, but rather simply make use of the implications of SED theory to develop of process for potential ZPE harvesting. Currently, only a semi-classical analysis using SED has been used to predict this shift of ground state orbital energies. Although much of SED theory has been applied successfully in producing results that are consistent with standard quantum mechanics [40,42,44], particularly with the inclusion of spin in the SED analysis [48], there have not been any reports to date in which this orbital energy shift has been replicated using quantum electrodynamics. An exploratory experiment to test for a shift in the molecular ground state of $H_2$ gas flowing through a 1 μm Casimir cavity was carried out, but without a definitive result [49].

### 2.3.2. The Extraction Process

In a 2008 patent [8], Haisch and Moddel describe a method to extract power from vacuum fluctuations that makes use of this ground state reduction inside Casimir cavities. The process of atoms flowing into and out from Casimir cavities is depicted in Figure 6. In the upper part of the loop, gas is pumped first through a region surrounded by a radiation absorber, and then through a Casimir cavity. As the atoms enter the Casimir cavity, their orbitals spin down and release electromagnetic radiation, depicted by the outward pointing arrows, that is extracted by the absorber. On exiting the cavity at the top left, the ambient ZPF re-energizes the orbitals, depicted by the small inward pointing arrows. The gas then flows through a pump and is re-circulated through the system. The system functions like a heat pump, pumping energy from an external source to a local absorber. The amount of power produced in a small toaster-sized system producing roughly $10^{22}$ transitions into and out from Casimir cavities per second was estimated to be approximately 1 kW. The power required to pump the gas through the cavities was estimated to be well below 1 W [8].

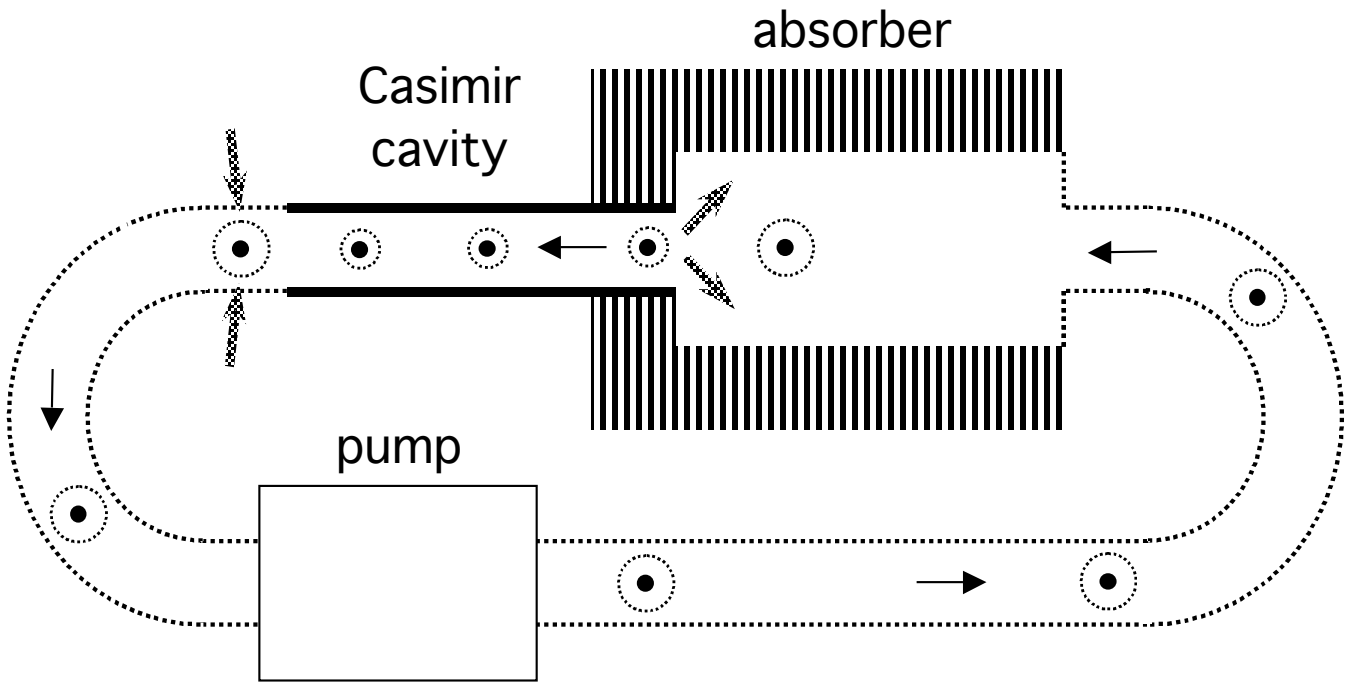

**Figure 6.** System to pump energy continuously from the vacuum, as proposed by Haisch and Moddel [8]. As gas enters the Casimir cavity the electron orbitals of the gas atoms spin down in energy, emitting Larmour radiation, shown as small arrows pointing outwards. The radiant energy is absorbed and extracted. When the atoms exit the Casimir cavity, the atomic orbitals are recharged to their initial level by the ambient zero-point field, shown by the inward pointing small arrows. The figure is reprinted from Ref. [49].

Initial studies on energy emission from this system have been carried out [50]. We first examine the results and then discuss the thermodynamics issues that are related to the process.

### 2.3.3. Experimental Test of Radiant Emission Due to Gas Flow

For a working gas of xenon, a Casimir cavity spacing of approximately 0.1 μm is required to suppress the frequency for the outer orbital [8]. In an initial set of experiments cavities were formed using spaced-apart optical flats, but because of the narrow spacing required, even the thick optical flats bowed too much to provide sufficiently uniform spacing. More reliable cavities were formed from Whatman polycarbonate Nuclepore flexible membranes with pore sizes of 0.1, 0.2, and 0.4 μm. To form metallic Casimir cavities gold was thermally evaporated at two angles onto the pore walls. The coverage was confirmed by scanning electron microscope (SEM) inspection, as shown in Figure 7.

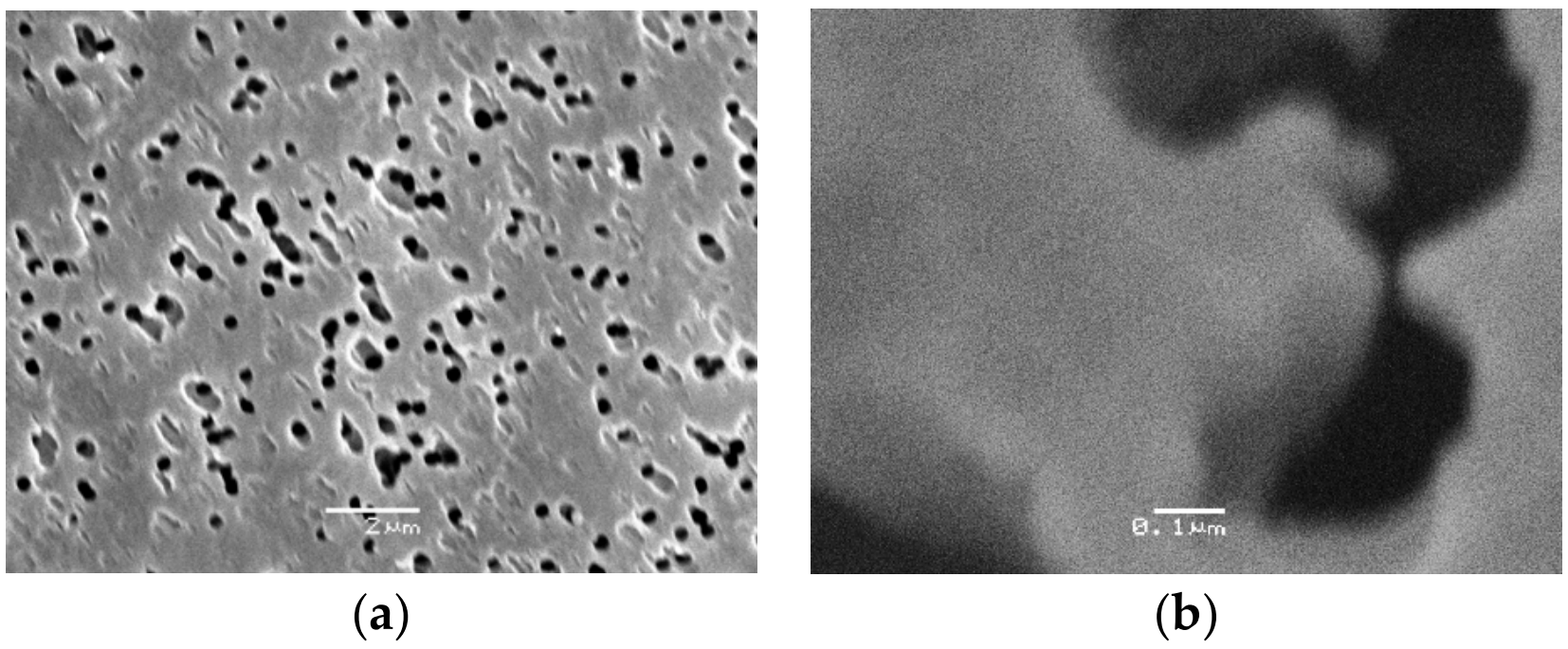

(**a**) (**b**)

**Figure 7.** SEM images of coated Nanopore polycarbonate membrane shown at two magnifications. In (**a**) the white line corresponds to a length of 2 μm, and in (**b**) to 0.1 μm. The figure is reprinted from Ref. [49].

Coated and uncoated membranes having a diameter of 25.4 mm were mounted into a holder and placed in an evacuated stainless-steel chamber. Gas flowed into the top of the membrane at a pressure of 1 to 10 Torr, and was pumped out below, with the pumping modulated at 0.5 Hz to facilitate lock-in detection of emission from the device. Above the device a pyroelectric detector measuring emission in the wavelength range of 0.6 to 5 μm was placed outside the chamber facing the device through an infrared-transparent ZrSe window. Additional experimental details are described in Ref. [49].

Four different gases (He, Ar, $N_2$, and Xe) were used to test both coated and uncoated membranes. Emitted radiation from filters with a 0.2 μm pore size is shown in Figure 8.

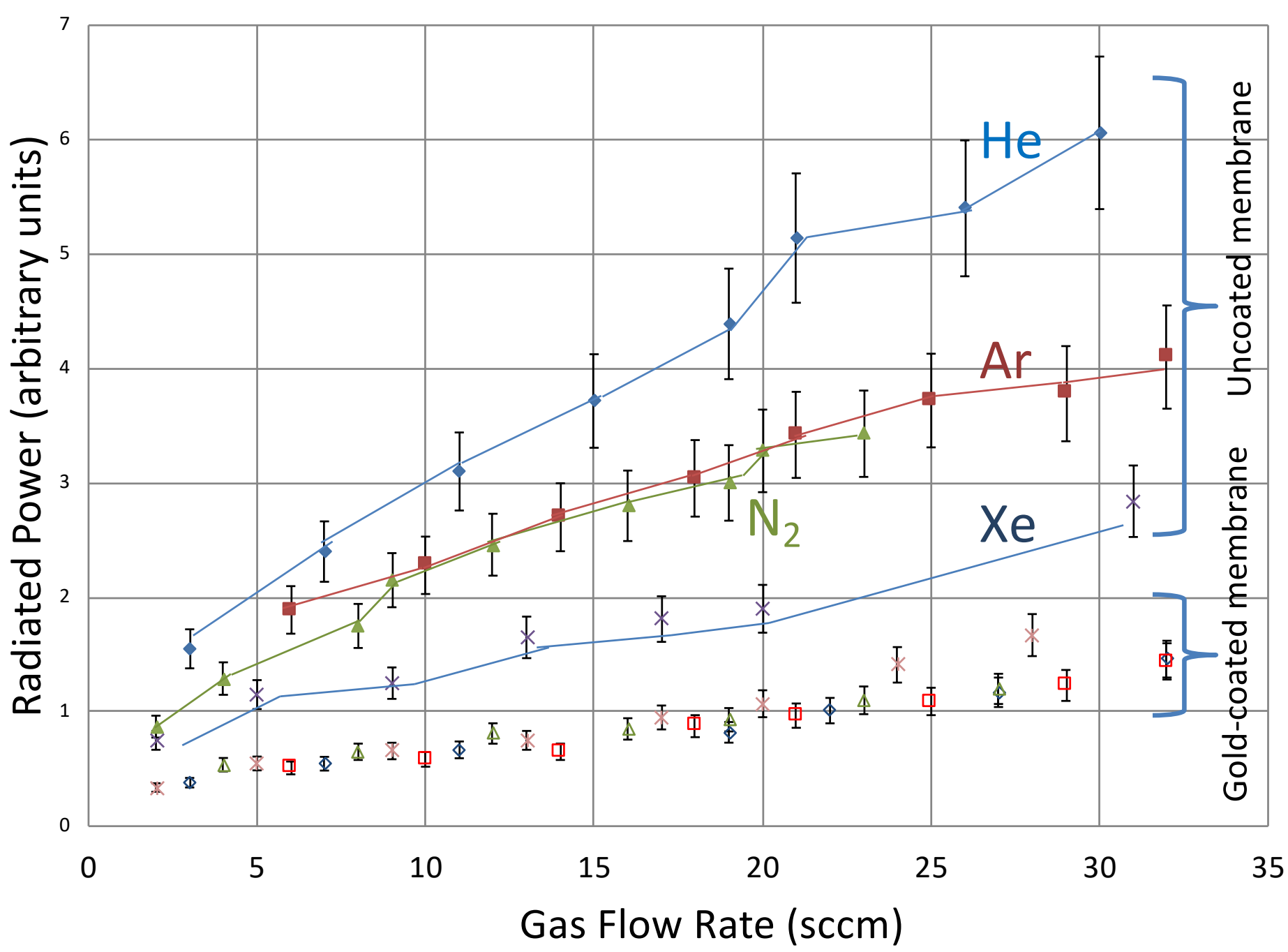

**Figure 8.** Measured radiation emitted using $N_2$, Ar, Xe and He gasses flowing through uncoated and gold coated polycarbonate filters having 0.2 μm pore size [49].

Emission of radiation from all the membranes was clearly measured, with lower output from the gold-coated membranes than from the uncoated polycarbonate. To assess how the radiation was produced a series of tests was carried out.

2.3.4. Test of Frictional Heating as a Source for the Observed Radiation

We compared the radiated power from the filters with different pore diameters. In Figure 9 we show the signal measured for helium gas flowing through 0.1, 0.2, and 0.4 μm pore diameter membranes. The highest signal was registered from the 0.1 μm filter. One possible explanation for why that diameter produced the largest signal is that the smaller sizes suppress ZPFs having frequencies that more closely match the atomic outer orbital frequencies, and hence, give rise to greater radiation from the ZPE. Alternatively, the smaller diameters produce a higher pressure drop and could give rise to more frictional heating, as discussed below.

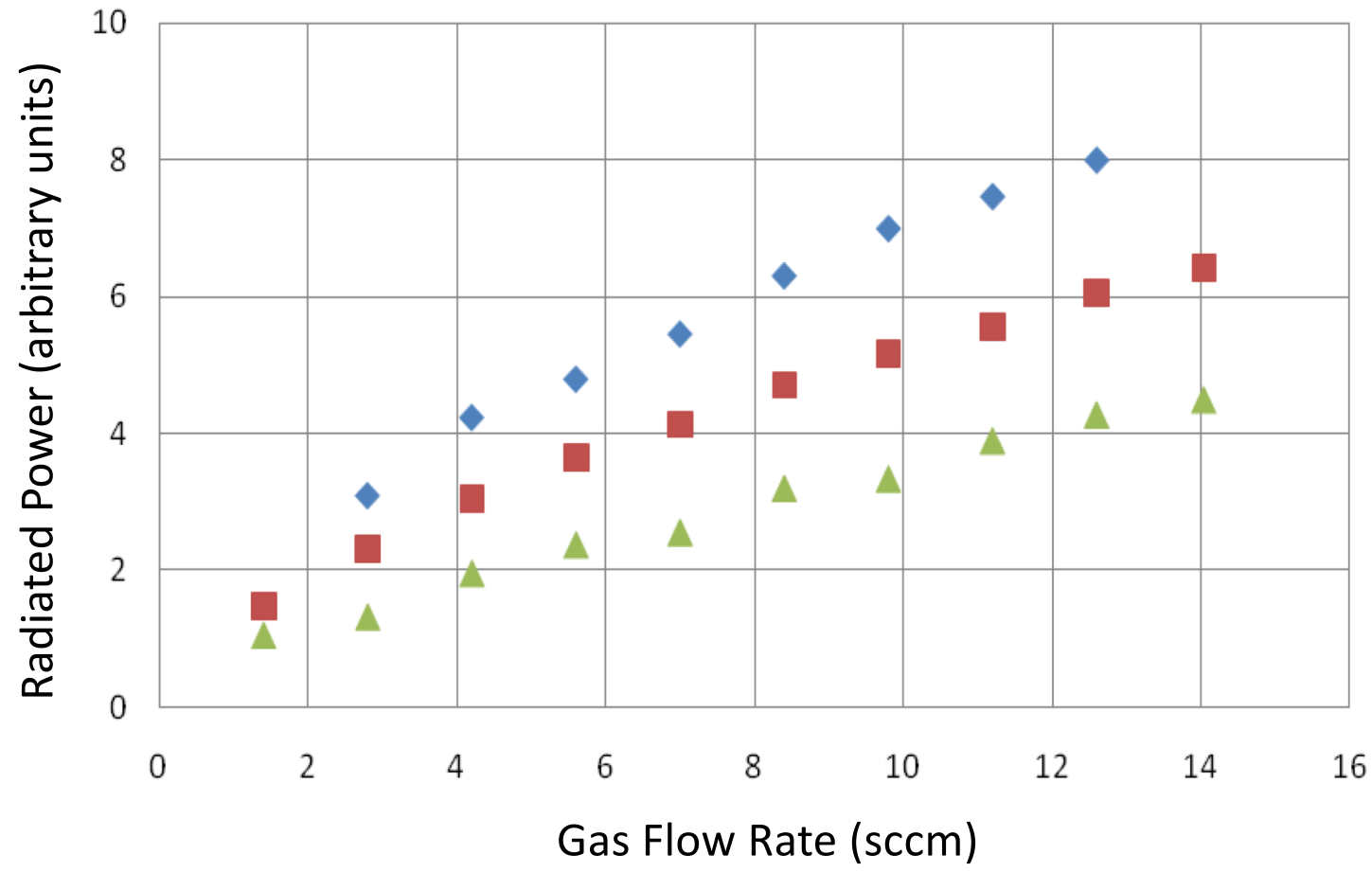

**Figure 9.** Effect of diameter on radiated power, as measured for He gas flowing through uncoated polycarbonate filters of 0.1 (blue diamonds), 0.2 (red squares), and 0.4 μm (greed triangles) pore diameter sizes.

To test for frictional heating, we examined whether the membrane might work as a passive element dissipating heat as a result of the pressure drop of gas flowing through it. Results for different gases flowing through 0.2 μm uncoated membrane are shown in Figure 10.

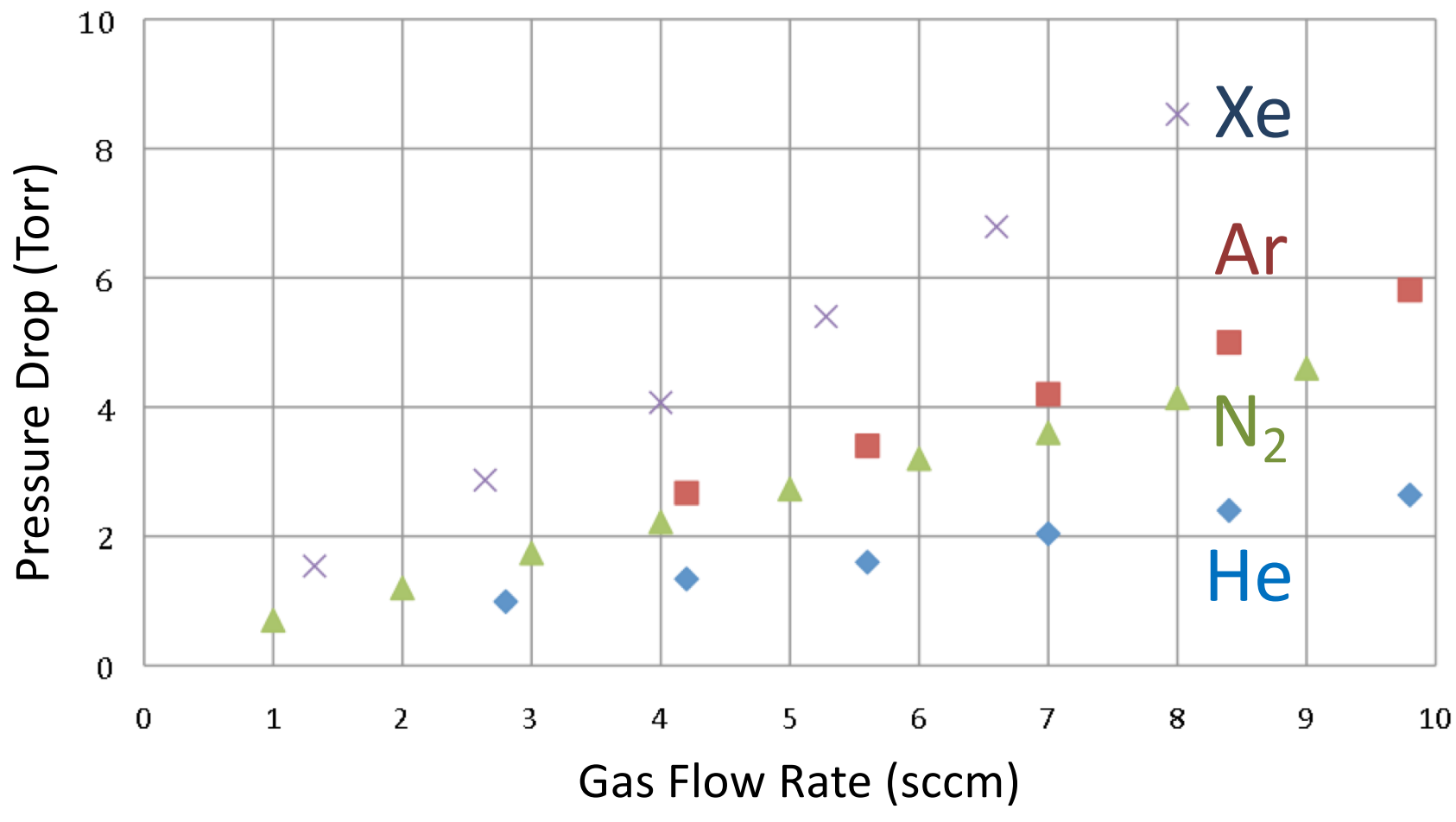

**Figure 10.** Pressure drop verses gas flow through 0.2 μm uncoated filter for different gases.

These results are in excellent agreement with the studies done by Roy and Raju on the gas flow through microchannels [51]. The pressure drop depends linearly on gas flow and the slope is different for different gases with xenon being the most resistive and helium being the least. If these frictional losses are what produce heating and the measured radiation, the power should depend quadratically on the gas flow. This would amplify the differences seen for the different gases, with xenon producing the largest heating and helium the smallest, as shown in Figure 11.

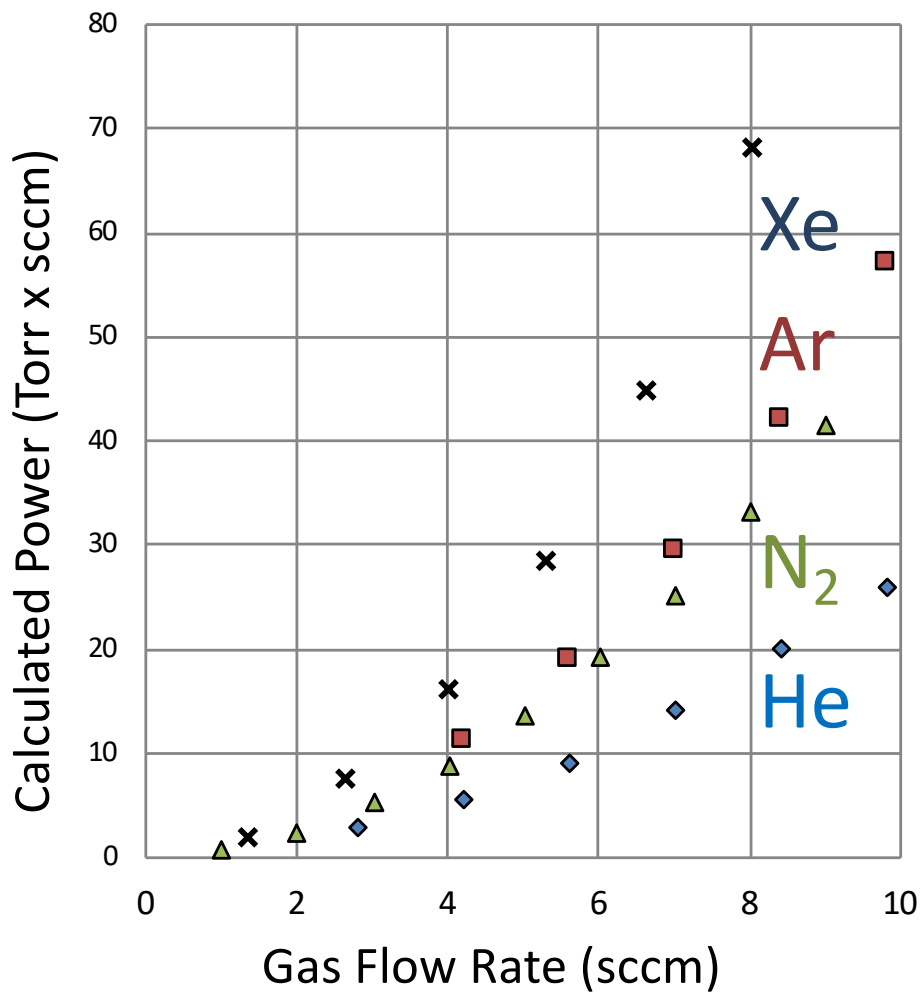

**Figure 11.** Calculated power from frictional loss based upon pressure drop for each of the gases.

This trend in radiated power is the opposite of what was observed in the data of Figure 8, calling the frictional heating model into question. On the other hand, helium is more thermally conductive than xenon, and so if the signal measured in Figure 8 somehow involves thermal conduction—despite the fact that it is radiation and not conduction that is being measured—the greater thermal conductivity of helium may give rise to gas trend that was observed.

Another issue with the frictional heating model is that the measured power, shown in Figure 9, depends linearly rather than quadratically on the flow. This observation does not necessarily rule out frictional losses as a source for the measured radiation. The reason is that the system has more elements taking part in the process, with gas in the chamber transferring its energy to the walls. This might affect the power dependence on the flow and cause it to appear linear. The frictional heating does not appear to be the source for the measured signal both because the gas order is wrong and because the dependence on pressure is wrong. Still, it cannot be ruled out entirely.

#### 2.3.5. Test of the Joule-Thomson Effect as a Source for the Observed Radiation

Another possible mechanism for the emission could be the Joule–Thomson effect [52]. To test for this, we replaced the membrane with a 12.5 μm thick Mylar film having a single hole 3 mm in diameter, far too large to give rise to any measurable ZPF effect. The results are shown in Figure 12. The phase measured by the lock-in amplifier was 180° off from the phase measured in the experiment with the nanopore membrane, indicating cooling. For xenon, argon, and nitrogen the Joule–Thompson coefficient is expected to produce cooling from expansion under the measurement conditions. We could clearly distinguish the heating we observed from flowing the gases through the Casimir cavities from this measured Joule–Thompson effect. This rules it out as a source for the observed radiation.

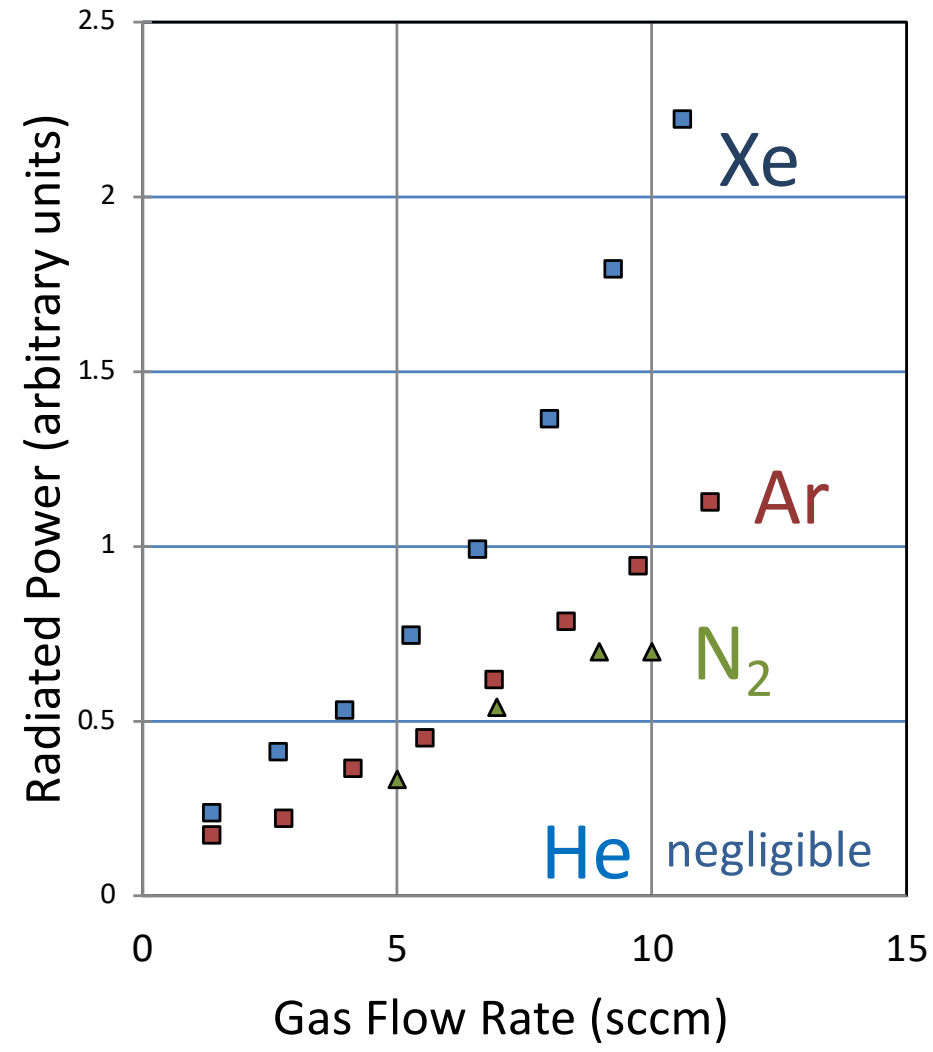

**Figure 12.** Test of Joule–Thomson effect. Emitted radiation was measured for $N_2$, Ar, Xe and He gasses flowing through a Mylar film having a 3 mm diameter hole. The phase of the lock-in signal indicated a cooling effect.

### 2.3.6. Turbulence as a Potential Source for the Observed Radiation

Turbulence is another possible heat source if the Reynolds number is sufficiently high, which would indicate that the flow is not dominated by viscosity [53]. In our experiment the rate at which the gas flowed through the nanopore cavities was below 2 cm/s, which results in a Reynolds number that is far below what would be required to cause turbulence at the exit. Furthermore, the viscosities of the gases used in the experiment have values that are very close to each other (Xe: 0.021 cP, Ar: 0.021 cP, $N_2$: 0.0166 cP, He: 0.0186 cP). Even if turbulence somehow did produce radiation from the membrane it could not explain the differences in signal levels for the different gases.

### 2.3.7. Absorption/Adsorption as a Potential Source for the Observed Radiation

When a gas is brought into contact with a solid surface its molecules are either absorbed into the solid or adsorbed on the surface, both processes being exothermic. We tested whether this could be a source for the observed radiation. Adsorption is proportional to the surface area. We used three different pore size membranes, 0.1 μm, 0.2 μm and 0.4 μm, having surface areas of 9.4 $cm^2$, 18 $cm^2$ and 12.6 $cm^2$, respectively. Signal from these different filters measured for a pressure of ~5 torr did not vary with surface area. Therefore, the absorption/adsorption model may be ruled out as a possible explanation for the observed radiation.

### 2.3.8. Expected Radiation Power

We can calculate the expected radiated power assuming that each atom releases 10% of its ground state orbital energy each time it enters a Casimir cavity, which would result in roughly 1 eV per transition [8]. For the flows we used that would result in the production of milliwatts of power. The measured power was below 1 μW for all the gases and samples. Assuming that roughly 20% of the emitted power was received by the detector facing the membrane, the measured power is three orders of magnitude smaller than expected. There could be multiple reasons for this discrepancy. An obvious one has to do with the pore shape, shown in Figure 7. It is far from uniform, so that only a small fraction of the gas atoms or molecules encountered Casimir cavities of the required dimensions. It is difficult to assess how much of a power deficit can be attributed to this, but it is clearly large. Another unknown has to do with the direction of the radiation. If there is radiation from the gas transitions, it may be directed to the walls of the Casimir cavity, causing them to heat up. In that case the radiation may be a secondary effect, with much of the power being lost to the environment by convection and conduction, and some radiation being outside the wavelength range of our detector.

If the power that was radiated does not come from ZPE, then it must come from the power supplied by pumping the gas through the cavities. Based on theory and experiments of gases flowing through nanopores [50] and consistent with the data and calculations of Figures 10 and 11, the power required to pump gas through the Casimir cavities is more than three orders of magnitude lower than the expected power that could be harvested [8]. Therefore, from a simple gas flow perspective the harvested power does not come from the gas pump.

The SED model adds an additional factor. One might argue that because the ground state energy of the gas atoms in the Casimir cavity is lower than that outside, they would have to surmount a potential hill to exit the cavity. The power to do that would have to come from the pumping. By that argument the power emitted when the gas entered the cavity would all be supplied by the pump, and no ZPE would be harvested. There is a flaw in that argument, however. The power would not have to be supplied by the pump because it is supplied by the ambient ZPF that the atoms encounter when they exit the cavity—that is the whole point of the system. Furthermore, we do not know of a mechanism that could translate the spherically symmetric energy radiation from the molecules into a back-pressure. Still, we cannot totally rule out this potential hill argument.

In addition to these possible loss mechanisms, there are other unknowns because there are not yet a precise theory and simulations that would enable us to predict what the desired dimensions should be. Given the ambiguities it is not yet possible to know whether the measured power is consistent with ZPE being the source.

#### 2.3.9. Deviations from Expected Results

The results deviated from what was expected in several ways.

1. The measured power is much lower than predicted, as described in the previous section. Some if not all this deviation can be attributed to inconsistent sizes and shapes of the nanopores.
2. Another unexpected result is that the uncoated polycarbonate membranes produced much more radiation than the gold-coated devices. It is expected that the metal-walled Casimir cavities are more effective than the dielectric-walled cavities in suppressing interior modes, although the latter does produce the Casimir effect [29] that are only slightly smaller [54]. A likely reason that more radiation was observed from the polycarbonate is that the emitted power heated the cavity walls and the emissivity of the polycarbonate walls and membrane is much greater than that of the gold.
3. We expected to see the greatest emission from the xenon atoms. Their outer orbital frequency corresponds to a wavelength (0.1 μm) that is suppressed in the 0.2 μm cavities (suppressed wavelength is ½ the cavity spacing). That suppressed wavelength is farthest from the wavelength corresponding to the helium orbital (0.05 μm). The opposite was observed. We do not know the reason for this, but it may have to do with more total energy being available from the helium atoms.

#### 2.3.10. Violations of the Second Law of Thermodynamics

Putting aside the fact that radiation was observed, we consider whether there are fundamental constraints on the process. The first question is whether it conflicts with the conservative nature of the Casimir force. It does not because although Casimir plates are used, they do not move as part of the process. Therefore, this process differs from the mechanical process described in Section 2.2.2, which does make use of cyclic Casimir plate motion in an attempt to extract power from Casimir attraction.

Next comes the question as to whether there is a detailed balance that would render the process invalid. If the gas were stationary, then we would expect a detailed balance of radiation to exist between the atoms and their environment at the entrance and exit to the cavity. However, the gas is flowing and in such a dynamic situation could circumvent the requirement for detailed balance. A key issue here are the time scales involved. If the gas enters the Casimir cavity on a time scale that is much slower than the equilibration time, then the entire process occurs in quasi-equilibrium and detailed balance applies. In such a case, there would be no excess energy to shed, e.g., to the absorber shown in Figure 6, because the orbitals would remain in equilibrium with the vacuum ZPF at each instant of the process.

In summary, although the process does not overtly violate the second law, it is not clear whether equilibrium considerations render it incapable of providing harvestable energy.

2.3.11. Future Work to Investigate Gas Flow through Casimir Cavities

A study based on the same concept of harvesting energy from ground state reduction was carried out by Henriques [55]. To avoid the detailed balance described above and to improve the chances for observing orbital energy shifts he optically excited the atoms as they entered the Casimir cavity. He was not able to detect any emitted radiation, but the reason might have been due to poor sensitivity in the detector.

Our results are tantalizing but unfortunately inconclusive. The investigation could be improved in several ways. One is to obtain nanopores having a more consistent size and shape so that they provide uniform cavities of the desired dimensions. This would allow for a better assessment of how much power can be emitted, and whether the emitted power exceeds the pumping power. Another would be to image the radiation to determine the location of its source, and to measure its spectral dependence to determine whether it is thermal and if so at what temperature.

## 3. Conclusions

The tremendous energy density in the zero-point field (ZPF) makes it tempting to attempt tapping it for power. Furthermore, the fact that these vacuum fluctuations may be distinguished from thermal fluctuations and are not under the usual thermal equilibrium make it tempting to try to skirt second law of thermodynamics constraints. However, the ZPF is in a state of true equilibrium, and the constraints that apply to equilibrium systems apply to it. In particular, any attempt to use nonlinear processes, such as with a diode, cannot harvest energy from a system in equilibrium. Detailed balance arguments apply.

The force exhibited between opposing plates of a Casimir cavity have led to attempts to make use of the potential energy to obtain power. This cannot succeed because the Casimir force is conservative. In any attempt to obtain power by cycling Casimir cavity spacing the energy gained in one part of the cycle must be paid back in another.

Thermal fluctuations, usually thought of as an expression of Planck's law, and ZPE vacuum fluctuations, usually associated with the ground state of a quantum system, are most often treated if they were separate forms of energy. However, the zero-point energy (ZPE) density given in Equation (1) may be re-expressed in the form [56]:

$$\rho(h\nu) = \frac{4\pi h\nu^3}{c^3} \coth\frac{h\nu}{2kT} \tag{2}$$

The fact that these two seemingly separate concepts can be merged into a single formalism suggests that thermal and ZPE fluctuations are connected fundamentally. More rigorously, Planck's law can be seen as a consequence of ZPE [57], and is "inherited" from it [58]. In more than a century of theory and experimentation we have not been able to extract usable energy from thermal fluctuations, and it might seem that we are destined to find ourselves in a similar situation with attempts to extract usable energy from ZPE.

There is, however, a distinction that can be drawn between the two cases, which has to do with the nature of the ZPE equilibrium state. The equilibrium ZPE energy density is a function of the local geometry. Two thermal reservoirs at different temperatures that are in contact with each other cannot be in equilibrium; heat will flow from one to the other. Two ZPE reservoirs having different energy densities that are in contact with each other can, however, be in equilibrium. For example, a Casimir cavity can be in direct contact (open at its edges) with the free space surrounding it such that the ZPE density inside and outside the cavity are different without any net flow of energy between the two regions. Furthermore, extracting ZPE from the vacuum does not violate the second law of thermodynamics [12].

Our apparent lack of clear success in extracting energy from the vacuum thus far leads to two possible conclusions. Either fundamental constraints beyond what have been discussed here and the nature of ZPE preclude extraction, or it is feasible and we just need to find a suitable technology.

**Author Contributions:** G.M. developed the analysis of the various zero-point energy harvesting methods. O.D. and G.M. conceived of and designed the experiment, and O.D. carried them out. G.M. wrote the paper.

**Funding:**

This work was funded by HUB Lab, Coolescence, and DARPA under SPAWAR Grant No. N66001-06-1-2026.

**Acknowledgments:** Many thanks to B. Haisch and M.A. Mohamed for stimulating discussions about vacuum energy, R. Cantwell and M. McConnell for help with the experiments, and to S. Grover, B. Haisch, B.L. Katzman and J. Maclay for comments on the manuscript.

**Conflicts of Interest:** G.M. owns stock in a company founded to develop the gas-flow energy harvesting technology, Jovion Corporation, but plays no active role in the company. Besides that, the authors declare no conflict of interest. The funding sponsors had no role in the design of the study; in the collection, analyses, or interpretation of data; in the writing of the manuscript, and in the decision to publish the results.